\documentclass[12pt,preprint]{aastex}
\shortauthors{Saripalli, Subrahmanyan, \& Udaya Shankar}
\shorttitle{Giant radio galaxy J0116$-$473}
\begin{document}

\title{A Case for Renewed Activity in the Giant Radio Galaxy J0116$-$473}

\author{Lakshmi Saripalli, Ravi Subrahmanyan}
\affil{Australia Telescope National Facility, CSIRO, 
Locked bag 194, Narrabri, NSW 2390, Australia}

\and

\author{N. Udaya Shankar}
\affil{Raman Research Institute, 
Sadashivanagar, Bangalore 560 080, India}

\begin{abstract}
We present ATCA radio observations of the giant radio 
galaxy J0116$-$473 at 12
and 22 cm wavelengths in total intensity and polarization. 
The images clearly reveal 
a bright  inner-double structure within more extended
edge-brightened lobe emission. The
lack of hotspots at the ends of the outer lobes, the strong core and 
the inner-double structure
with its edge-brightened morphology lead us to suggest that this giant radio
galaxy is undergoing a renewed nuclear activity: J0116$-$473
appears to be a striking example of a radio galaxy where a
young double source is evolving within older lobe material.
We also report the detection of a Mpc-long linear feature which is
oriented perpendicular to the radio axis and has
a high fractional polarization.
\end{abstract}


\section{Introduction}

The concept of episodic activity in radio galaxies, with
each phase manifesting itself as an
extended radio structure, was inherent in the models suggested for  
sources with X-shaped structures and powerful radio galaxies with 
wings \citep{leahy84}.  Restarting beams
following an interruption in nuclear activity was again suggested as
a cause for source structures which appeared to have partial jets
\citep{bridle86}. This 
idea gained support from the observations of 3C388 by \citet{roettiger94} 
in which the lobe spectral index distribution revealed two distinct 
regions. Observational
indications coupled with simulations of the development of extended
radio structures have suggested that episodic activity may play an
important role in the evolution of at least some categories
of radio sources \citep{baum90,clarke91}.

In a study of the 
morphologies of a sample of giant radio galaxies, \citet{subrahmanyan96} 
drew attention to a variety of morphological features which were suggestive of 
interrupted nuclear activity. Recently, the WENSS discovered several 
giant radio galaxies exhibiting double-double morphologies which have been 
attributed to renewed nuclear activity \citep{schoenmakers00b} and the study
is indicative of a higher incidence rate of such inner doubles among large
radio galaxies. As a consequence, systematic studies of the role of
episodic nuclear activity are possible for the category of giant radio 
galaxies. The 
relatively long timescales ($\sim 10^{8}$ yr; Komissarov and Gubanov 1994)
over which radio lobes remain visible 
after the central activity which energizes them stops makes synchrotron lobes 
useful indicators of any past activity phases in radio galaxies.  

Such studies have interesting implications for the fuelling of the 
central engine and the conditions 
under which a renewal of nuclear activity may occur.
Moreover, these studies may
address the question of the role of such recurrent activity in the
attainment of the extraordinary sizes in the giant radio sources.

J0116$-$473 was previously imaged as part of a study of the morphologies
in radio galaxies of megaparsec dimensions \citep{subrahmanyan96}.
This object was noted as exhibiting unusual characteristics: lack of hotspots 
in a source which had properties consistent with FR-{\sc II}
type \citep{fanaroff74} radio galaxies 
and an elongated structure extending along a direction perpendicular to the
jets. It was hypothesized that the morphological features in this
giant radio galaxy --- and indeed some others in the sample --- might be
a manifestation of recurrent nuclear activity.  \citet{subrahmanyan96}
concluded that the large
sizes of giant radio galaxies may be a result of a restarting of their central
engines in multiple phases of activity along roughly similar directions. 

We are currently following up on our earlier hypothesis with case
studies of the giant sources which showed evidence of recurrence in activity. 
In this paper, we present higher dynamic range Australia Telescope Compact
Array (ATCA; see The Australia Telescope 1992) images of 
J0116$-$473  in total intensity and polarization. The ATCA observations 
presented here provide new evidence for a restarting of activity in this
giant source.  In the next section we describe our observations.
In later sections we discuss the role of recurrence in the creation of the
unusual morphological features in this source. We postpone comparison of 
the source features and parameters with  other sources 
exhibiting similar inner double structures to a later paper.

\section{Observations and Imaging}

J0116$-$473 has an angular size of about 12 arcmin and it 
was observed with ATCA at 12 and 22 cm wavelengths
in several array configurations in order to
image a range of angular-scale structures from a few arcsec to 
several arcmin.  Observations were made during 1999 January to April
in an extended 6.0C 6-km array, a 1.5C 1.5-km array, a 750C 750 m array 
and a compact 375-m array.  Full Earth-rotation synthesis observations 
over 12 hr were made in each of the configurations in order to 
cover the visibility domain with the East-West array.  22 and 12-cm
wavelength visibilities were recorded simultaneously in two bands
of 128 MHz each centered at 1384 and 2496 MHz.  Full polarization
measurements were recorded in order to image the source in Stokes
I, Q and U.  The continuum band was covered in 16 independent channels.

The flux density scale was bootstrapped to B1934$-$638 whose flux density
was adopted to be 14.9 and 11.14 Jy respectively at 1384 and 2496 MHz.
The data were calibrated, imaged and deconvolved using standard
procedures and the {\sc miriad} software routines.  
To avoid bandwidth
smearing effects in the wide-field images made with the wide continuum 
bands,  images were constructed
using the bandpass-calibrated multi-frequency channel data 
adopting bandwidth synthesis techniques.   Images were deconvolved
using the \citet{clark80} algorithm with the \citet{cornwell83} modification
to suppress CLEAN instabilities.  At least two iterations of phase 
self-calibration were performed; the visibility amplitudes were
not self-calibrated.

\section{The Source Components and Physical Parameters}

In Figure 1 we show a total intensity image of J0116$-$473 at 22 cm made
with a beam of FWHM $10.2 \times 9.1$ arcsec$^{2}$. The giant radio galaxy is 
seen to have (i) two large, diffuse, edge-brightened lobes which do not show
any hotspots at the ends, and (ii) what appears to be another double
source, smaller in size and sharing the same radio core, whose lobes have
a much higher brightness.  A peculiar elongated structure is also seen
extending nearly a megaparsec in a direction perpendicular to the radio axis
and intersecting the radio axis just south of the core.  The source is
seen to have a similar morphology at 12 cm.

J0116$-$473 has a total flux density of 2.9 Jy at 1376 MHz and 1.6 Jy
at 2496 MHz.  The overall spectral index is 0.93 
(we define the spectral index
$\alpha$ by the relation: $S_{\nu} \sim \nu^{-\alpha}$) between
these two frequencies. The source is at a redshift of 0.146 \citep{danziger78}
and the total radio luminosity at 1376 MHz is 
$2.2 \times 10^{26}$ W Hz$^{-1}$ (herein we adopt cosmological parameters
H$_{\circ}$ = 65 km s$^{-1}$ Mpc$^{-1}$, $\Omega_{m}$ = 0.3  and
$\Omega_{\Lambda}$ = 0.7).  The largest angular size is 12.5 arcmin
and we infer the linear size to be 2.1 Mpc.

The radio images detect a radio core with a flux density of 11 mJy
at 1376 MHz and with a flat spectral index, computed between 1376 and
2496, of $-0.1$.

\subsection{Outer Lobes of J0116$-$473}

The total radio power of J0116$-$473 places it in the category
of powerful radio galaxies; however, the outer lobes lack compact 
hotspots.  In their place, we can recognise only diffuse warm-spots
towards the lobe ends. 

Both the outer diffuse lobes (hereinafter we refer to the northern and southern
outer lobes as N1 and S1, the corresponding inner lobes are called N2 and S2)
appear relaxed although they are seen to be 
well bounded as inferred from the bunching of the contours at the edges over 
much of the lobes.  The sharp outer boundaries are indicative of a 
confinement of the synchrotron plasma by the external intergalactic
medium (IGM).

The intensity distribution of linearly polarized emission at 22 cm,
with a beam of 12 arcsec FWHM, is shown in Figure 2 as a greyscale
representation with contours overlayed. The electric field polarization
vectors are also overlayed, with the vector lengths proportional to the
fractional polarization.  
From the orientations of the polarization
vectors at 12 and 22 cm, we have derived the distribution of
rotation measure (RM) over the source: values of RM
are small and have a mean of 3 rad m$^{-2}$ with a 1-$\sigma$ spread
of 4 rad m$^{-2}$ (see Figure 3).
It may be noted here that the observed difference in the 
orientations of the polarization vectors at 12 and 22 cm are small;  
therefore, we have assumed that our RM values
inferred from 2-frequency polarization images do not suffer from
$n \pi$ ambiguities. The vector orientations in Figure 2
have been corrected for the derived Faraday rotation.

The fractional polarization is enhanced along 
most of the outer boundaries of both lobes and the projected magnetic 
field lines are circumferentially oriented along these boundaries: the 
polarization properties
are suggestive of an ordering of the field owing to a 
compression of the lobe plasma along the outer boundaries. Fractional
polarization in the range 20-30 per cent are measured over 
most of the two outer lobes with 
values exceeding 50 per cent along the lobe boundaries. 
We observe significantly more 
structure in both the outer lobes in polarized intensity as compared to
that in total intensity. The dips in polarized intensity
in the vicinity of the intensely polarized regions
in N1 and S1 are likely to be due to beam depolarization
because the E-vectors on either side of these dips show
large differences in their position angles.  

The image of polarized intensity at 12 cm is very similar to that at 22 cm;
we computed the distribution of depolarization ratio over 
the source as a ratio of the fractional polarizations at 22 and 12 cm.
No regions of the source appear to have noticable depolarization at a 
resoution of 12 arcsec and we also do not see any depolarization
asymmetry: on the average, the fractional polarization at 22 cm
is within 5 per cent of that at 12 cm in both lobes.
The lack of evidence for internal Faraday rotation and depolarization 
asymmetry in J0116$-$473 is 
consistent with similar findings for other GRGs \citep{willis90,lara99}.  

It may be noted here that the morphology in the northern lobe 
appears to indicate that the diffuse warm-spot region at its outer end was 
probably fed  by a flow which 
entered the lobe along the axis defined by the inner double and
bent by almost $90^{\circ}$ at the north-eastern boundary
of the lobe. The situation may have been similar to that in sources
which show `rim hotspots': these  
radio galaxies, which are several times smaller, have hotspots that are
seen along the rim of a lobe and the
jet is thought to enter along the 
rim and bend through a large angle to create a terminal hotspot
at its end ({\it e.g.}, 3C135, 3C192; Leahy et al. 1997). It may be that we
are viewing a relict of a rim-type hotspot complex in this giant
radio galaxy and we identify the warm spot as the position where the 
beam may have terminated. 

The projected magnetic field distribution inferred from Figure 2 
may provide a clue as to how
the beam propagated in the past. 
In the north-east regions of the lobe, the B-field vectors are 
circumferential and would be parallel to a beam which
propagated into the northern 
lobe along its eastern rim and curved to terminate at the location of
the peak in polarized and total intensity.  
A sharp flip in the position angle of the E-vectors is seen 
at the location of the peak
in going from NE to SW across the diffuse lobe: if 
the south west of the lobe is made of post-hotspot material, this would
imply that the field here is oriented perpendicular to the flow.
An interesting feature is a
loop-like distribution of the projected B-field over the lobe with reduced 
polarized intensity in the central regions towards the southern end.

A de-collimated flow, as inferred for the northern lobe, is 
uncharacteristic of 
powerful radio galaxies where narrow beams usually feed hotspots at the lobe 
ends; large-opening-angle beams are more typical of low-power sources whose
lobes are edge darkened.  It is interesting to ask whether a decrease in
the beam power, as the central engine activity in a powerful radio source 
ceases, is accompanied by a transition to an FR-{\sc I} type flow.

We observe a large-scale structure in the RM distribution over the
source: as seen in Figure 3, the RM is low in a band 
running along the length of the radio galaxy at a position angle 
of about $-21^{\circ}$ and with a width of 2 arcmin. 
In Figure 4 we display the profile of this RM variation: the 
image shown in Figure 3 was rotated counter clockwise through $21^{\circ}$ 
and binned (averaged) along declination to construct the profile plot. 
The profile clearly shows that the mean RM is close to zero in the 2 arcmin
wide band and values on 
either side of the band average around 5 rad m$^{-2}$.

In sensitive observations of high Galactic latitude
fields,  \citet{wieringa93} and later \citet{haverkorn00} 
reported detection of 
linear structures in polarized intensity which showed abrupt changes in
RM along directions transverse to the filaments.
These bands have widths and derived RM values
similar to the feature we observe in J0116$-$473. 
However, as seen in Figure 3 where contours of total intensity
are overlayed on the RM image, there appears to be a correspondence
between the RM and total intensity distributions: in N1 and S1 the
brightness shows a step decrement at the location of the band with low RM.
Moreover, the band with low RM has
a position angle similar to the source axis although it is not
symmetric with respect to the radio core position.
These indicate that the RM structure may be intrinsic to the source
and not a foreground galactic feature.  If such an interpretation
were correct, it may imply a spatial separation of 
entrained thermal plasma in the lobes: the low RM channel may trace
relicts of the beam plasma from past activity.
It may be noted here that there is no corresponding feature in 
the distribution of depolarization ratio; however, this is understandable
owing to the low values of RM. 

In Figure 5 we show the distribution of spectral
index, computed between 22 and 12 cm and using images with 
20 arcsec FWHM beams, over the entire source.  
Both outer lobes show a steepening of the 
spectral index away from the lobe ends: this is a characteristic feature of 
the lobes in powerful radio galaxies and has, traditionally, been interpreted 
as owing to spectral aging in the cocoon plasma as it backflows from hotspots
towards the central core.
However, in the case of J0116$-$473, there are no hotspots at the ends;
therefore, we hypothesize that 
the outer lobes are probably relicts, of a powerful radio galaxy, in which the
beams from the central engine no longer terminate.

The optical field of J0116$-$473 does not indicate evidence
for significant galaxy overdensity in the vicinity of the radio source
\citep{subrahmanyan96}.
Because the diffuse outer lobes are located
far from the host galaxy and in an environment which is poor in galaxy
density, we may expect a relatively low IGM pressure.
\citet{subrahmanyan93} used $COBE$ constraints on the
Comptonization of the CMBR spectrum and concluded that giant radio galaxies
were not thermally confined by the IGM.
Using current limits on the Comptonization $y$-parameter 
of $\mid y \mid < 15 \times 10^{-6}$ \citep{fixen96},
and assuming that the uniform IGM gas was ionized by redshift $z=5$,
we expect any uniform IGM to have a present day thermal pressure
at most $10^{-17}$ dyne cm$^{-2}$.  The outer lobes
N1 and S1 have peak brightness of about 10 mJy beam$^{-1}$ in 22 cm images
made with 10 arcsec FWHM beams and the synchrotron emission has a spectral
index $\alpha \sim 1$; assuming that the lobes have a line-of-sight
path length of 0.5 Mpc and making standard minimum energy assumptions
\citep{miley80}, we infer that the lobe synchrotron plasma has 
an energy density of at least $3 \times 10^{-13}$ erg cm$^{-3}$.  
Clearly, the lobes are overpressured with respect to the 
ambient IGM and, consistent with the
indications from the total intensity and polarization observations
discussed above, the lobes --- although relicts ---  are 
likely expanding and ram-pressure confined by the IGM.  

The two outer lobes 
are somewhat dissimilar in structure: 
N1 is seen to have a more spherical shape
where as S1 appears cylindrical or elongated. There is a
conspicuous lack of emission between N1 and the core 
(except for the sharply
delineated N2) and this is in contrast to the southern outer lobe S1
in which the lobe emission appears to extend to the core.

\subsection{The Inner Double}

A higher resolution image of the inner double, made at 12 cm 
and with a beam 
of $4.4 \times 4.1$ arcsec$^{2}$ FWHM, is shown in Figure 6.
The total flux density of the inner double, excluding 
the core, is 0.26 and 0.17 Jy respectively at 22 and 12 cm; the inner lobes 
have an overall spectral index of 0.7 which is significantly 
flatter than that of the outer lobes.  The radio luminosity of the
inner lobes is $2 \times 10^{25}$ W Hz$^{-1}$ at 1376 MHz: the 
inner double, by itself, has a luminosity which is on the dividing
line between FR-{\sc I} and FR-{\sc II} sources.

The inner double, in isolation, appears like the two 
lobes of a radio galaxy.  
The two components are both edge brightened and collinear with the
core.  Both are sharply bounded towards their leading edges with decreasing
surface brightness away from their ends. The northern lobe (N2) of the
inner double is more diffuse, without hotspots and is relatively short in 
length. In contrast, the southern inner lobe (S2) is elongated and has a 
series of emission peaks along its length: these peaks are not
all collinear.  In spite of the striking morphological differences between the
northern and southern components of the inner double, the ends of the 
components
are equidistant from the core: the inner double has arm lengths
which differ by less than 10 per cent and the overall linear size of the
inner double is 600 kpc. 

We have detected a one-sided jet 
close to the core and directed towards S2: 
the jet appears to be directed to the northern-most peak in S2.  
The jet is
much narrower than the lobe and emission peaks which constitute S2. 
The core, lobe N2, the jet and the two peaks in lobe S2 
which are closest to the core are all well 
aligned. However, two sharp bends are seen to occur in S2 at the locations of 
the middle two peaks in the lobe. 
The jet is unresolved in our 12 cm image 
made with a beam of 4.2 arcsec FWHM (Figure 6) 
and we estimate the jet width to be $< 8$ kpc; in contrast, the inner lobes
are well resolved and have transverse widths of about 20 arcsec.
The detection of the narrow jet is a strong reason to believe that S2 and
N2 are indeed lobes of a restarted activity and that they are not
visible parts of jets transporting energy to the outer lobes S1 and N1.
This view is supported by the increasing transverse widths of N2 and S2 
towards the core and the symmetric locations of the outermost peaks in N2 and 
S2 on either side of the core.

If the inner double is a new episode of activity that is
propagating within the older lobe plasma, the ambient medium for the inner
and outer doubles would be different and differences in their
interactions with their respective ambient media might be reflected in
their spectral and polarization characteristics. 
The properties of the inner double and its evolution 
may then be used as a probe of the older lobe plasma in which it
is embedded. In J0116$-$473, S2 is surrounded by emission from S1 and this
provides an opportunity for examining the evolution of the inner lobes
within the outer cocoon material. 

The inner double and outer lobes share
the same radio core and the axis of the inner double appears aligned with
the axis of beams which fed the outer lobes: we would expect that S2
is embedded in the cocoon of S1 and that N2 is either propagating
in the IGM or a low-surface-brightness southern extension of N1.
If we assume cylindrical symmetry for S1, we may expect that S2 would
be embedded in S1 even if the axes of the outer and inner doubles
were misaligned in 3-dimensional space.
The lobe separation symmetry observed in N2 and S2 also argues against
large angles between the axis of the inner double and the plane of the sky.

The polarization in the emission from the 
inner double at 12 cm is shown in Figure 7 with a beam of 4 arcsec FWHM.
Care was taken
to avoid contamination from the diffuse emission surrounding S2: only 
visibilities with spatial frequencies exceeding 3 k$\lambda$ were used in 
making the image.  In N2 and S2, the projected B-field vectors 
appear to follow the total intensity contours at the 
leading edges of the lobes: a property commonly seen 
in FR-{\sc II} type radio galaxies.  There is also evidence for a 
circumferential magnetic field along the edges of the lobes
away from the ends: this is similar to the behaviour seen in 
the outer double and other FR-{\sc II} type radio galaxies. 
The fractional polarization is enhanced all along the edges of
N2 and along the eastern edge of S2 and has values exceeding
50 per cent in these regions; the depolarization ratio computed between
22 and 12 cm over the inner lobes is unity within the errors.
The low polarized intensity (and fractional polarization) in the western 
edge of S2 may be a consequence of beam depolarization because the E-vectors
sharply change orientation from being circumferential along the edge to
being perpendicular to the source axis away from the edge. 
In the remainder of the inner lobes, the fractional
polarization is about 20-30 per cent.  The projected B-field
away from the lobes edges is oriented perpendicular to the 
source axis; however, the jet detected close to the core has a projected 
B-field configuration which is oriented along the jet axis: the
arrangement is common in jets in powerful radio galaxies.
The lack of depolarization asymmetry lends additional support for
the inner and outer lobes both being close to the plane of the sky.

The spectral index distribution over the inner lobes was computed from
images at 2496 and 1376 MHz --- made using visibilities 
in the restricted common range 1.7-30 k$\lambda$ --- which were
convolved to a final beam of $7 \times 5.5$ arcsec$^{2}$ at 
p.a. of $0^{\circ}$.  The spectral
index image is featureless at this resolution and has a value 
about 0.7 over most of N2 and S2. The spectral index image was binned
along RA and the profile of the average spectral index in
declination is shown in Figure 8.  There does not appear to be any
significant spectral index gradient along either N2 or S2; if anything, 
there is marginal indication for a flattening of the spectral 
index toward the core in both the lobes.  The jet emerging from the core
toward S2 has a spectral index of 0.6.

Both N2 and S2 show uniform RM values close to zero (Figure 3). 
In Figure 9, we show the RM distribution in the vicinity of S2 
where it is seen that the lobe appears to be surrounded by regions
which have a distinctly different value of RM. 
The correlation in the spatial distribution
of RM with total intensity points to the possibility that 
the RM structure is intrinsic to the source. It is interesting that the
higher RM values avoid the area covered by S2, indicating greater
entrained thermal plasma and/or ordered fields in the ambient cocoon 
close to the edge of the advancing inner lobe.

To summarize, the inner double has an edge-brightened morphology, a 
distinctive feature of powerful FR II-type 
sources; however the absence of spectral gradients in the inner 
lobes is uncharacteristic of double radio sources. If spectral gradients
along lobes of double radio sources represent age of synchrotron plasma,
the absence of significant gradients in the inner lobes of J0116$-$473 suggests
that the advance speed of the ends of the source is unusually high compared 
to sources growing in any thermal IGM. This is consistent with the suggestion
that lobes advance within relict cocoons with speeds 0.2-0.3$c$
\citep{schoenmakers00a} which is much greater than the advance speeds of
0.03$c$ inferred for powerful radio galaxies \citep{scheuer95}.

In the restarted jet model of \citet{clarke91} the new jets are overdense 
with respect to their ambient medium and propagate almost ballistically 
unable to form hotspots. In these conditions circumferential fields and 
high degrees of polarization are not expected to be seen towards the 
ends of the restarted jets. In giant radio galaxies, however, 
\citet{kaiser00} argue that the longer timescales involved
allow for sufficient entrainment of ambient material across the
contact discontinuity and into the older cocoon plasma and this
raises the thermal density to levels adequate for the 
formation of hotspots at the ends of the new jets.
The circumferential B-field, accompanied by high degrees of 
polarization, seen around the leading ends of N2 and S2 
constitute the typical signature expected of compressed synchrotron plasma. 
The observations are indicative of an evolution for the ends of the
inner lobes that is similar to that for the  outer lobes and 
in powerful radio galaxies where light jets form
hotspots on meeting with a denser ambient medium. 
The advance of a restarted jet is also predicted to result in
a bow shock structure within the older cocoon plasma with an intensity
contrast same as that of structure at its head. \citep{clarke92}.  Our 
observations do not reveal any evidence for such a bow shock feature
in total intensity; similar to the case in
other giant radio galaxies with inner double structures \citep{kaiser00}.

\section{Recurrent activity in J0116$-$473}

The observational evidence presented in the preceeding sections 
lead us to postulate the following scenario for the formation of the radio 
structure: the outer lobes were formed in an earlier phase of activity in 
which powerful beams from the central engine formed hotspots at the ends
of the source and the edge-brightened lobes. 
That phase of nuclear activity ceased and the beams feeding the hotspots
and lobes discontinued.  Since then, the
hotspots have relaxed and expanded into the warm spots and the
beam channels were pinched off.  The outer lobes N1 and S1 
are the visible relicts of the past activity phase.  The beams
were recently re-activated following the interruption and these new beams 
have formed the inner lobes N2 and S2.
The position angle of the inner double is seen to be the same as
the axis of the beams that formed the outer double and this suggests that 
the re-activated central engine maintained the axis of ejection. 
It may be noted here that the warm spot in the northern lobe is probably not
along the axis of the past activity phase (see section 3.1).  

We see that while there are no glaring separation asymmetries in this radio
galaxy for the outer and inner double structures, morphologically there are
clear side to side asymmetries and an indication that the asymmetry  repeats
over the two activity epochs.  Such a similarity in the morphologies of the
inner and outer components was also reported in at least two
other giant radio galaxies 
with inner double structures (B1450+333 and 1834+620; 
\citet{schoenmakers00a}). S2 has a cylindrical shape 
similar to the southern outer lobe S1; N2 appears more
relaxed and more spherical as is the case for the northern outer lobe N1.
If one views the inner double as a stage
through which the outer double might have evolved, 
the lobe morphology would appear to have been set early in the 
evolution. It would be interesting to study the role of the bar-like feature
in creating medium differences close to the host galaxy that might cause
the side-to-side asymmetries; this would require the linear 
feature to have existed before the formation of the outer lobes. 

The SuperCOSMOS digitization of the UKST red plate 
OR 18623 has been used to extract the optical image of the host galaxy and is
shown in Figure 10. Contours of the smoothed image (using a 2-arcsec FWHM 
Gaussian) have been overlayed after subtracting the mean sky background. 
The cross marks the location of the radio peak as determined from a 4-arcsec 
resolution image of the core at 2496 MHz.
There is seen to be a concentration of objects within 100 kpc of the host 
elliptical galaxy. One object located 15 arcsec to the north west, 
at a distance of about 50 kpc, is relatively bright
and is seen in the red image to be clearly extended towards --- and
within the extended envelope of --- the radio galaxy host, suggestive of a
possible interaction. Optical observations are suggested to substantiate
this as well as to look for any evidence for episodicity in the host galaxy
properties which might be related to multiple nuclear activity in this galaxy. 

\subsection{Activity Related Timescales}

From the lack of any hotspots in N1 and S1 we can obtain an estimate of
the time elapsed since the central engine activity stopped.
Assuming a maximum jet bulk flow velocity of $c$,  
the travel time of the jet material from the core to the presumed 
locations of the past hotspots in the outer lobes is at least 
$3 \times 10^{6}$ yr. 
A 1-kpc size hotspot of relativistic plasma may be expected to
disappear in a sound-crossing timescale of about $10^{4}$ yr. Because
hotspots are not seen in N1 and S1, we estimate that beams may have
stopped injecting energy into N1 and S1 at least $10^{4}$ yr ago.
 
Subsequently, if the outer lobes are overpressured with respect
to the ambient IGM (see section 3.1), N1 and S1 would have continued
to experience expansion losses.  Assuming 
that the ambient IGM has density at most 
$10 \Omega_{B} = 10 \times 0.019 h^{-2}$ \citep{burles99},
the outer lobes, which have internal energy density exceeding 
$3 \times 10^{-13}$ erg cm$^{-3}$, would have to expand at 
speeds at least $0.012 c$ in order to be ram-pressure confined.  
Expansion by factor $f$ would reduce the
luminosity by factor $f^{-(4\alpha+2)}$ \citep{leahy91a}: J0116$-$473 is 
currently seen
as a powerful radio source and if we assume that the 22-cm radio luminosity 
of the lobes was at most $10^{28}$ W Hz$^{-1}$
when the energy injection was switched off, 
we conclude that the relict lobes are
younger than $7 \times 10^{7}$ yr.

The previous activity phase may have stopped at most
$7 \times 10^{7}$ yr ago and, if the inner 
lobes advance into the relict cocoon with 
speed about 0.2-0.3$c$ \citep{schoenmakers00a}, we estimate that the
current activity phase commenced $3-5 \times 10^{6}$ yr ago.

\section{The Bar-like Feature}

In the southern lobe, we recognize an unusual bar-like 
feature just south of the core. The bar is sharply bounded 
along its northern edge and appears to be a distinct component.  
It has a length of at least 1 Mpc 
and is oriented nearly perpendicular to the source axis. 
As seen in Figure 2, this feature is the most
highly polarized region in the radio galaxy and appears to have a fractional
polarization exceeding 50 per cent. The projected magnetic field in the bar
is aligned along its length.
The radio spectral index of the bar emission,
as computed using our ATCA images at 12 and 22 cm,
is steep with $\alpha > 1.3$.

Examples of lobes of radio galaxies abruptly cutting off in sharp, straight
edges facing the parent galaxy have been cited in several radio galaxies
\citep{black92,gk00}. The lack of emission
along a columnar region about the core has been speculated as arising due to
the docking of the backflowing lobe plasma by an extended disk of 
cold gas, situated at the center of the host galaxy, 
with its axis along the radio axis.  These edges, seen previously
in normal-sized powerful radio galaxies, have extents $\leq 100$ kpc.
In the case of J0116$-$473, only one lobe 
reveals a sharp, straight edge towards the core and
it is seen to be nearly a megaparsec in length.  

Radio galaxies having such extended features at large angles to the main
source axis are usually 
recognised as winged and X-shaped sources. These extended,
fainter features are thought to be remnants of older activity or cavities
in the IGM --- excavated during an older activity --- into which newer lobe 
plasma has flowed \citep{leahy84}.
These sources are considered to have undergone a change in the direction of 
ejection over a large angle.
If we conjecture that the linear bar-like feature is a remnant of 
past activity (prior to the formation of S1 and N1) in the same 
host galaxy, the galaxy has since moved $\sim 100$ kpc
and this may take $\sim 10^{8}$ yr assuming velocities typical
for galaxies in small groups \citep{hickson97}.  
A relict with this age may not be 
expected to be visible owing to expansion and radiation losses; however,
backflow from S1 may render the cavity visible by replenishing it. 

The bar in J0116$-$473
has a high fractional polarization and B-field oriented along its length.
These properties are also seen
in low-surface-brightness wings of X-shaped sources (e.g. 0828+32,
3C223.1, 3C403 and 4C12.03); however, they are more 
striking in J0116$-$473.   If backflow from the 
southern outer lobe of J0116$-$473 flows into the relict cavity --- as
is required to render it visible ---  the uniform projected B-field
and the high fractional polarization
might be indicating an ordered flow along the bar and possibly 
a stretching of the fields as the plasma expands at the end of the
relict channel.

The radio structures seen in J0116$-$473 are not unlike those in 
the X-shaped radio galaxy 4C12.03 \citep{leahy91b}.
In this latter object, as well, there are three distinct structures which
may be associated with three successive activity phases:
an extended double along the main axis of the radio source, an inner
double along the same axis and a fainter bar-like feature with
straight edges of almost 800 kpc extent which makes a large angle
with the main axis of the radio galaxy.

In the scenario proffered above for the formation of the 
structures in J0116$-$473, the required large change in the ejection axis
between the first episode (responsible for the bar) 
and the second (responsible for N1 and S1) may be 
produced by an interaction with a sufficiently massive intruder that re-aligns
the black hole axis \citep{natarajan00}. 
The lack of change in the ejection axis in the restarting episode,
which resulted in the inner lobes N2 and S2, may be a result of 
an interaction involving a smaller intruder
that does not bring with it sufficient angular momentum to perturb the 
ejection axis significantly.  Alternately, constancy of the ejection axis
in restarting episodes might be a result of multiple encounters 
between the host galaxy with a single intruder \citep{schoenmakers00b}, 
in which the successive encounters alter the activity state of the
central engine without significantly perturbing the black hole axis.

\section{Summary}

We have presented 12 and 22 cm total-intensity and polarization ATCA
observations of the giant radio galaxy
J0116$-$473. The observations were carried out with the purpose of following up
on intriguing aspects noted for this galaxy in an earlier work.  
Our new higher dynamic range 
observations have provided much support for our earlier hypothesis of 
interrupted nuclear activity in this source. The inner double structure 
located within the much larger diffuse lobes is argued to be a pair of 
new lobes formed as a result of renewed activity in the core. The observations
show this inner double in detail revealing its edge brightened 
morphology, its symmetric location about the core, and a narrow jet. 

The 1-Mpc long bar-like feature close to the core is seen to be highly
polarized with fractional polarization as high as 50 per cent all along its 
length and with projected magnetic field vectors oriented along its length. 
We discuss a 
possible origin for this feature, suggesting it to be a result of earlier 
activity. Additionally, we note the presence of two other
unusual features seen in this source, a band of low rotation measure along the 
length of the source and a step in the rotation measure situated towards
the southern 
inner lobe. We briefly discuss possible causes for these features.

\acknowledgments

The Australia Telescope is funded by the Commonwealth of Australia for
operation as a National Facility managed by CSIRO. 
We thank the referee for useful suggestions
which led to investigation of aspects not presented in the original
version. We acknowledge the use of SuperCOSMOS, an advanced 
photographic plate digitising machine at the Royal Observatory of Edinburough,
in obtaining the digitised image of J0116$-$473 presented in the paper.

\clearpage

\begin{figure}
\epsscale{0.9}
\plotone{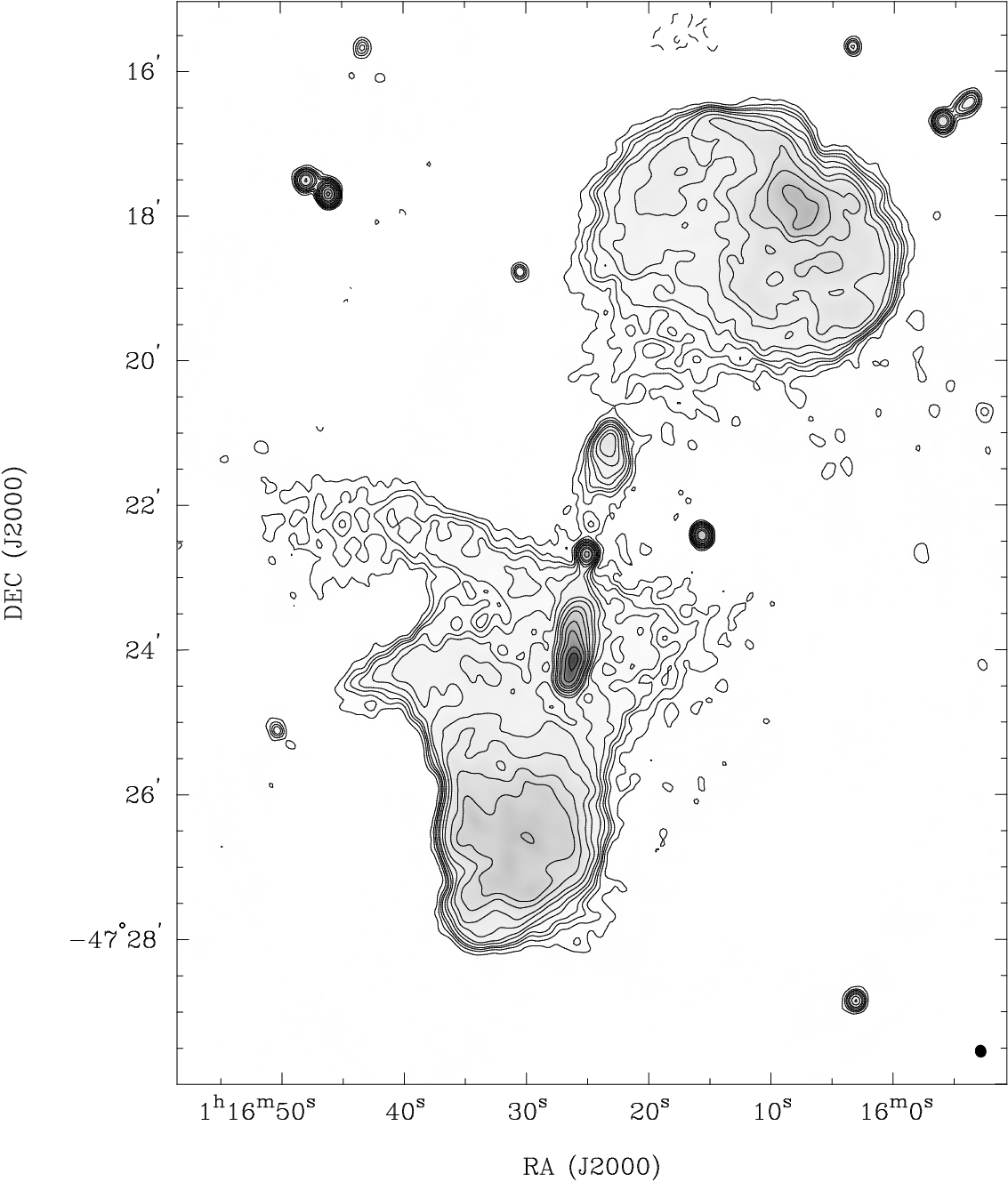}
\caption{
	J0116$-$473 at 1376 MHz made with a beam $10.2 \times 9.1$ arcsec$^2$
	at a position angle (p.a.) of $11^{\circ}$.  In this and in all
	following radio images, the FWHM size of the
	synthesized beam is indicated by the filled ellipse at the bottom
	right corner of the image; all images have been corrected for 
	the attenuation due to the primary beam.
	Contours are at 0.2 mJy beam$^{-1}$
	$\times$ ($-$1, 1, 2, 3, 4, 6, 8, 12, 16, 24, 32, 48, 64, 96, 
	128). \label{fig1}
	}
\end{figure}

\clearpage

\begin{figure}
\epsscale{0.9}
\plotone{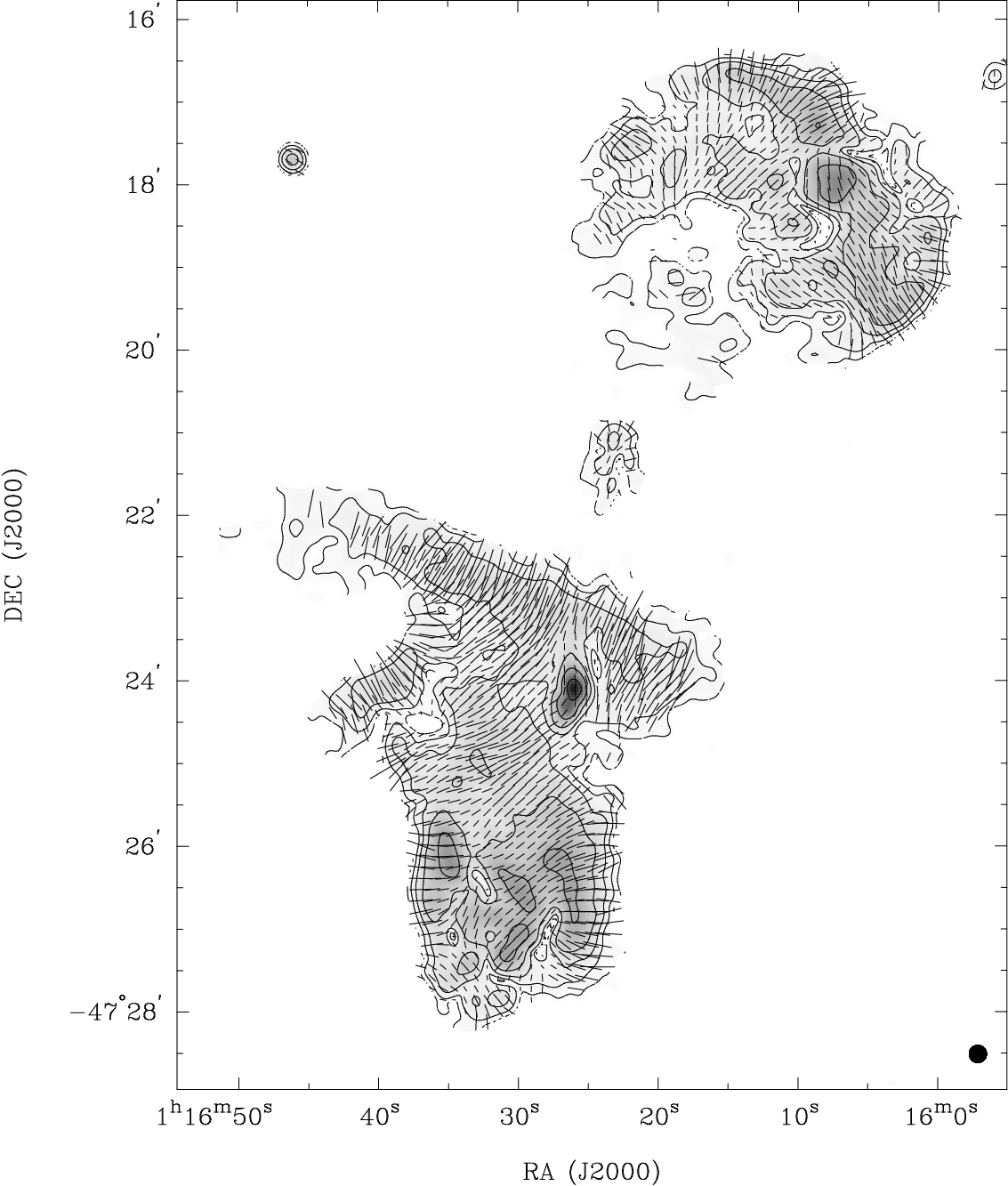}
\caption{
	Linear polarization of the 22 cm radio continuum from J0116$-$473
	at a resolution of 12 arcsec.
	The grey-scale and contour images are of the intensity of the
	linear polarization: contours at 0.2 mJy beam$^{-1}$ $\times$ 
	(1, 2, 4, 8, 16, 32).  The electric field vectors are
	displayed; the vector lengths represent fractional polarization
	using a scale of 2.7 per cent = 1 arcsec. \label{fig2}
	}
\end{figure}

\clearpage

\begin{figure}
\epsscale{1.0}
\plotone{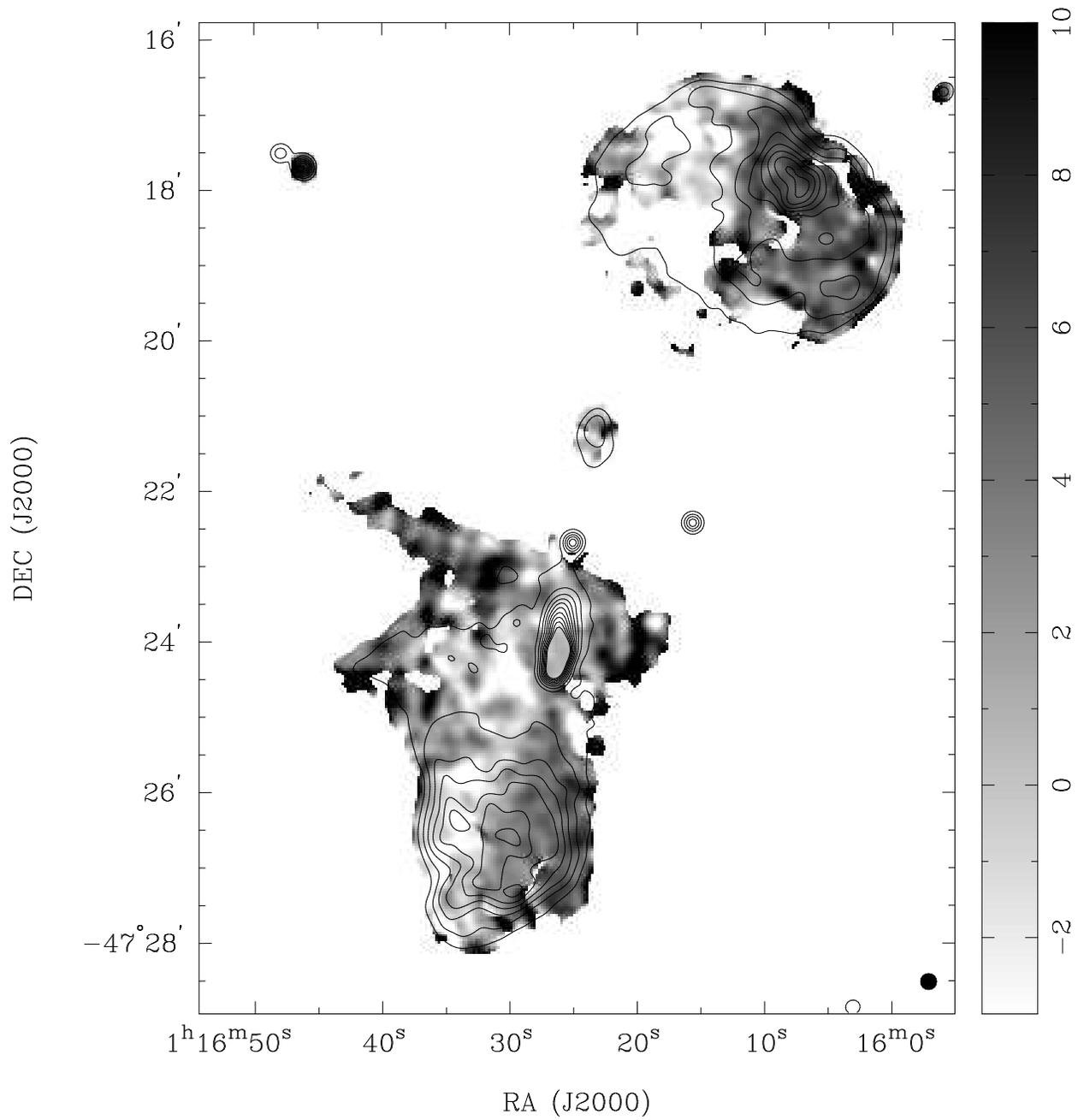}
\caption{
	Distribution of rotation measure (RM) over J0116$-$473.  The
	RM is computed using 12 arcsec FWHM resolution images of the
	polarization at 2496 and 1376 MHz, and is shown using grey-scales
	in the range $-3$ to 10 rad m$^{-2}$.  Contours representing
	the total intensity image 
	at 1376 MHz, with 12 arcsec FWHM resolution, are overlayed; contours
	are at 2 mJy beam$^{-1}$ $\times$ (0.5, 1, 2, 3, 4, 5, 6, 7, 8).
	\label{fig3}
	}
\end{figure}

\clearpage

\begin{figure}
\epsscale{1.0}
\plotone{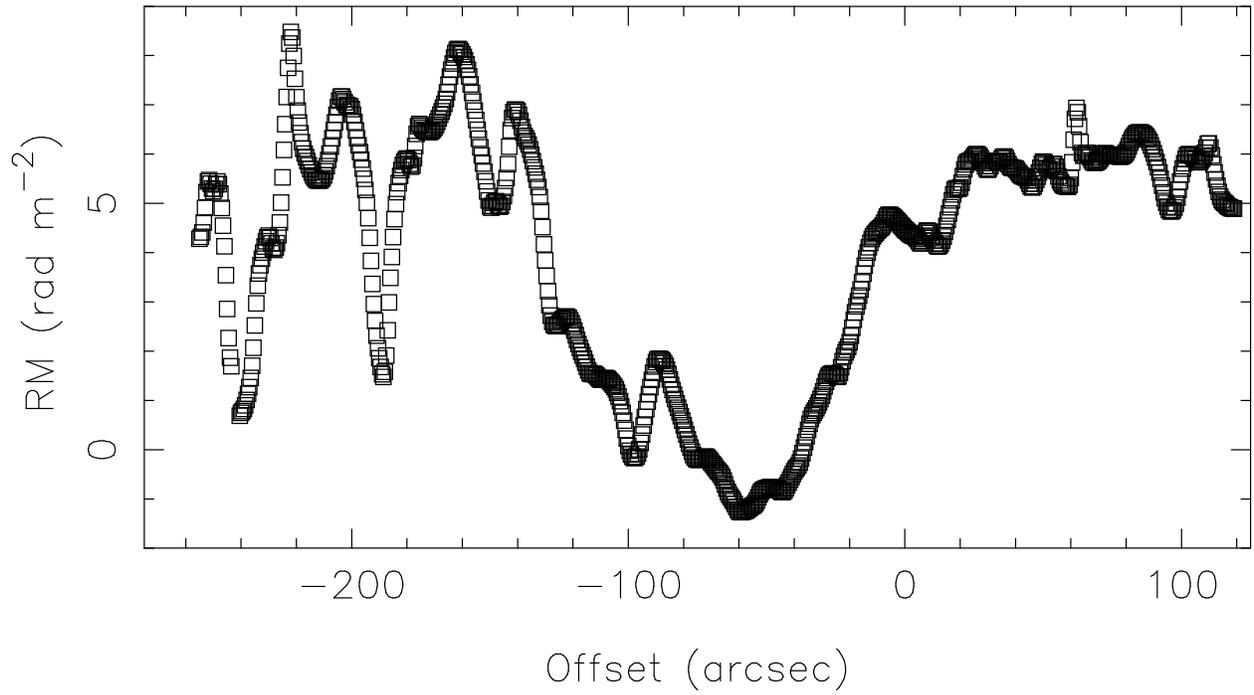}
\caption{
	The image represents the variation in RM across the source,
	computed along a position
	angle of $-111^{\circ}$, as a function of offset from
	the position RA: $01^{h}16^{m}20^{s}$, DEC: $-47^{\circ}22^{\prime}
	30^{\prime\prime}$.
	The image shown in Figure 3 was
	rotated counter clockwise through $21^{\circ}$ and binned (averaged)
	along declination to construct this profile plot.
	\label{fig4}
	}
\end{figure}

\clearpage

\begin{figure}
\epsscale{1.0}
\plotone{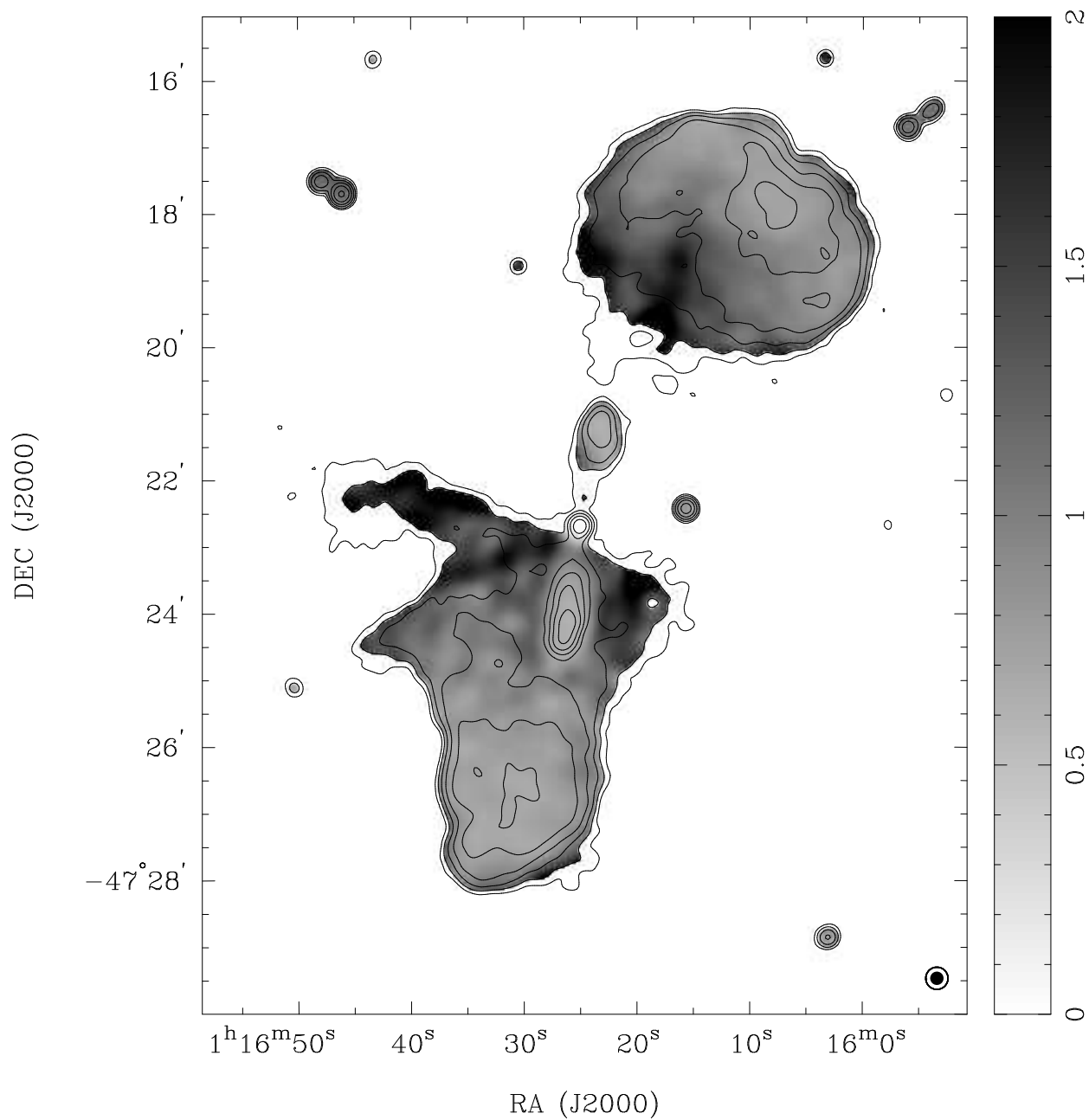}
\caption{
	Spectral index distribution across 
	J0116$-$473 computed between 2496 and 1376 MHz using images
	made with 20 arcsec FWHM beams.  
	The spectral index $\alpha$ is shown using grey scales 
	in the range 0 to 2 and
	is defined as $S_{\nu} \sim \nu^{-\alpha}$, where $S_{\nu}$ is the
	flux density at frequency $\nu$.  Contours represent 12 cm total
	intensity with beam 12 arcsec FWHM: contours
	are at 0.4 mJy beam$^{-1}$ $\times$ (1, 2, 4, 8, 16, 32). 
	\label{fig5}
	}
\end{figure}

\clearpage

\begin{figure}
\epsscale{0.80}
\plotone{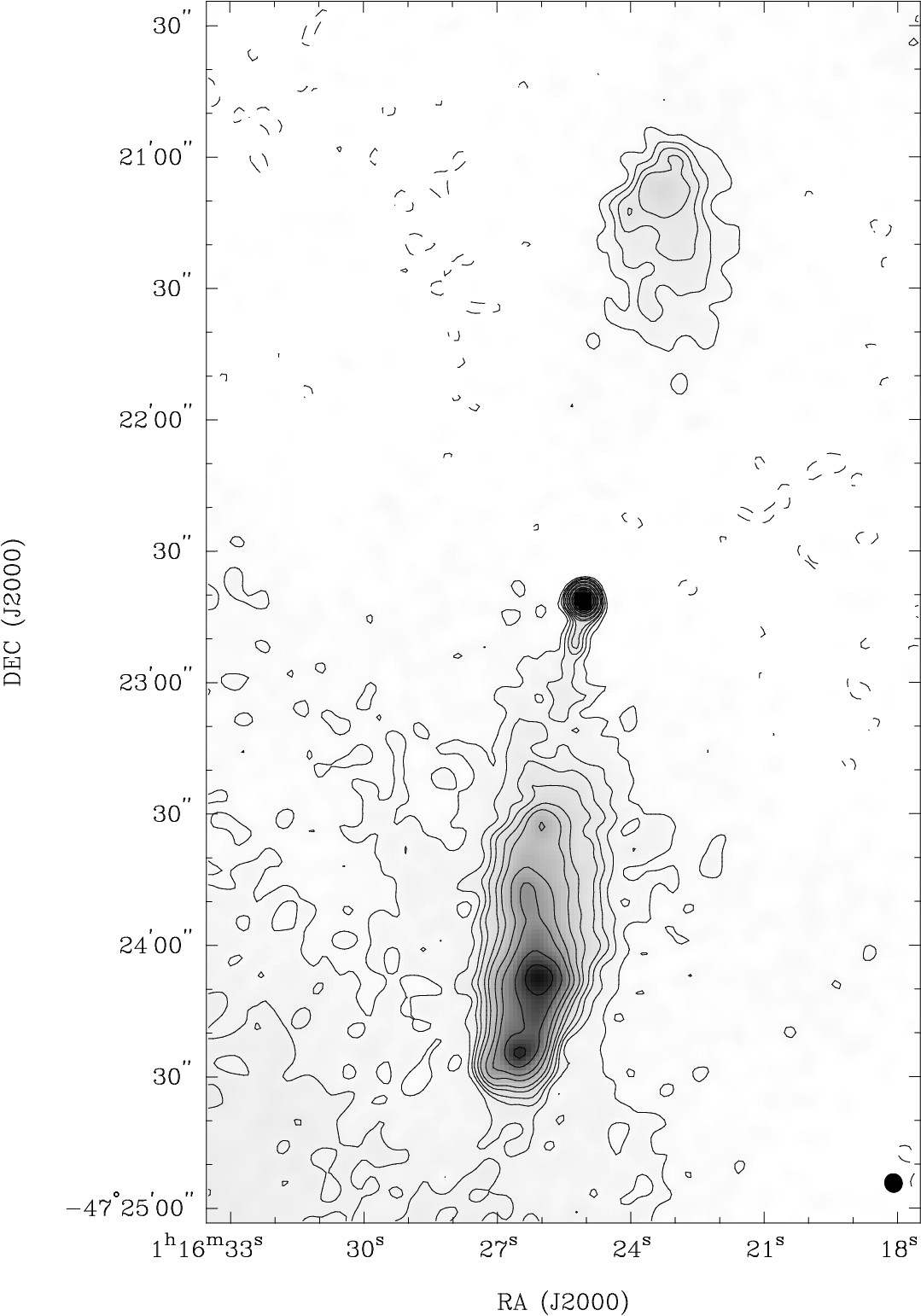}
\caption{
	Radio image of the inner lobes of
	J0116$-$473 at 2496 MHz made with a beam $4.4 \times 4.1$ arcsec$^2$
	at a p.a. of $15^{\circ}$. The image is displayed using contours
        and grey-scales;
	contours are at 0.15 mJy beam$^{-1}$
	$\times$ ($-$1, 1, 2, 3, 4, 6, 8, 12, 16, 24, 32, 48, 64, 96, 
	128).  \label{fig6}
	}
\end{figure}

\clearpage

\begin{figure}
\epsscale{0.85}
\plotone{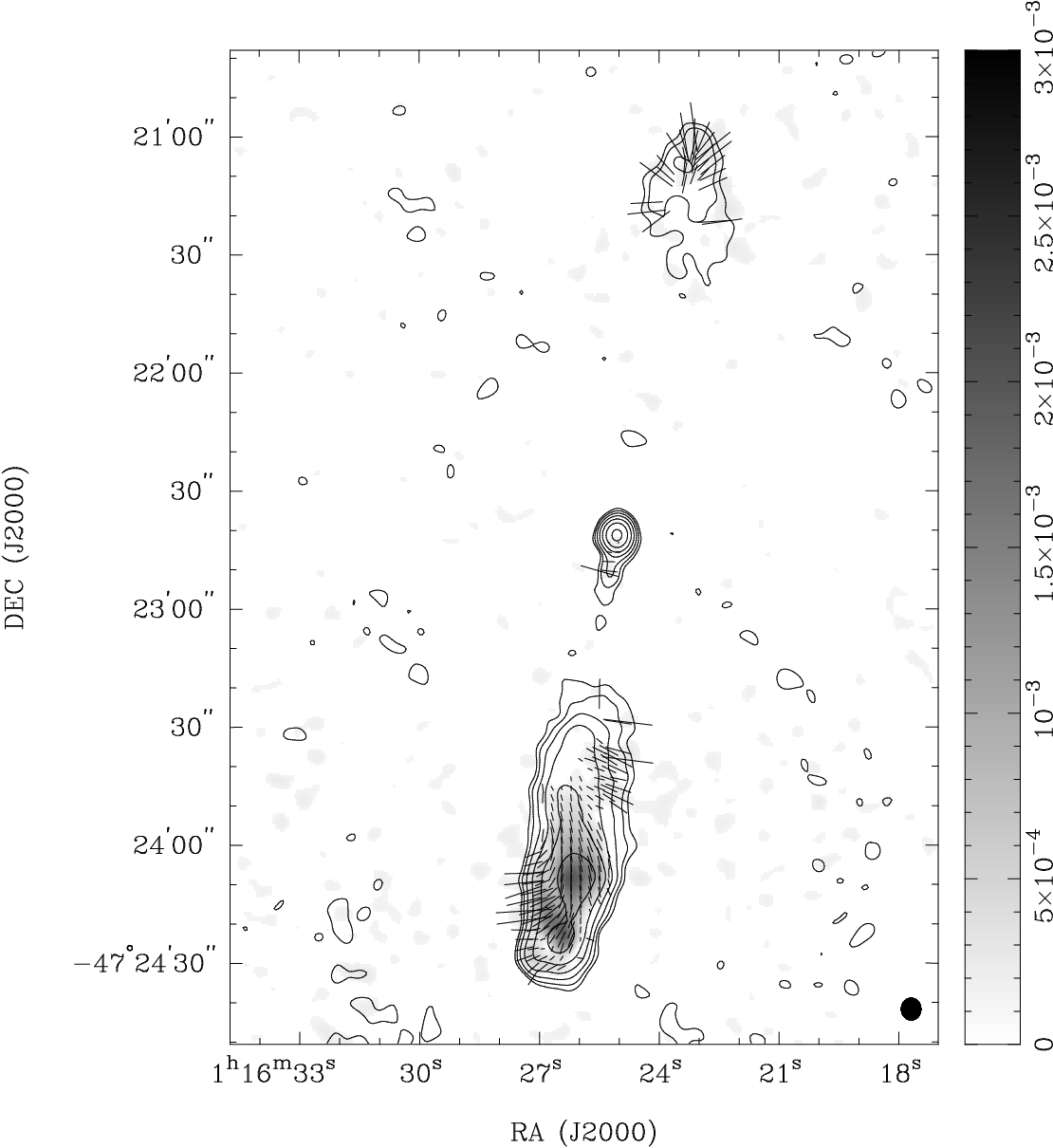}
\caption{
	Polarization of the 2496 MHz emission from the inner lobes
	at a resolution of 4 arcsec FWHM computed from images of the
	source made using only visibilities exceeding 3k$\lambda$ wavelengths
	(so that contributions from the diffuse emission are absent).
	The polarized intensity, in Jy, 
	is shown using grey-scales.  Contours of total intensity are
	overlayed; contours at 0.15 mJy beam$^{-1}$
	$\times$ (1, 2, 4, 8, 16, 32, 64).
	Electric field vector orientations are
	shown with length representing the fractional polarized
	intensity using a scale: 1 arcsec = 13.4 percent.
	\label{fig7}
	}
\end{figure}

\clearpage

\begin{figure}
\epsscale{1.0}
\plotone{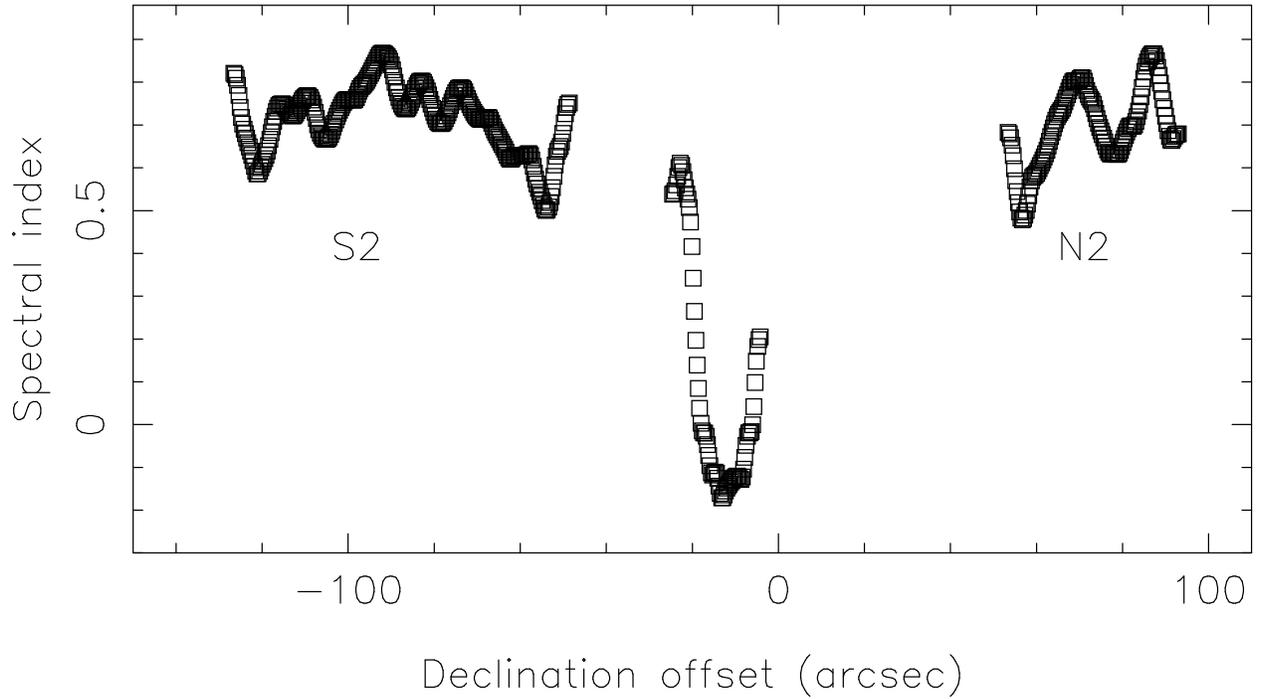}
\caption{
	Spectral index distribution along the inner lobes of
	J0116$-$473.  Images of the inner lobes
	at 2496 and 1376 MHz were made 
	using visibilities in the range 1.7-30 k$\lambda$
	and convolved to a final beam of $7 \times 5.5$ arcsec$^{2}$ at 
	p.a. of $0^{\circ}$.  These images were used to compute the 
        spectral index distribution which was binned
	along RA to obtain the profile of the average spectral index along
	declination: the profile is displayed versus offset from 
	DEC: $-47^{\circ}22^{\prime}30^{\prime\prime}$.	 
	\label{fig8}
	}
\end{figure}

\clearpage

\begin{figure}
\epsscale{0.85}
\plotone{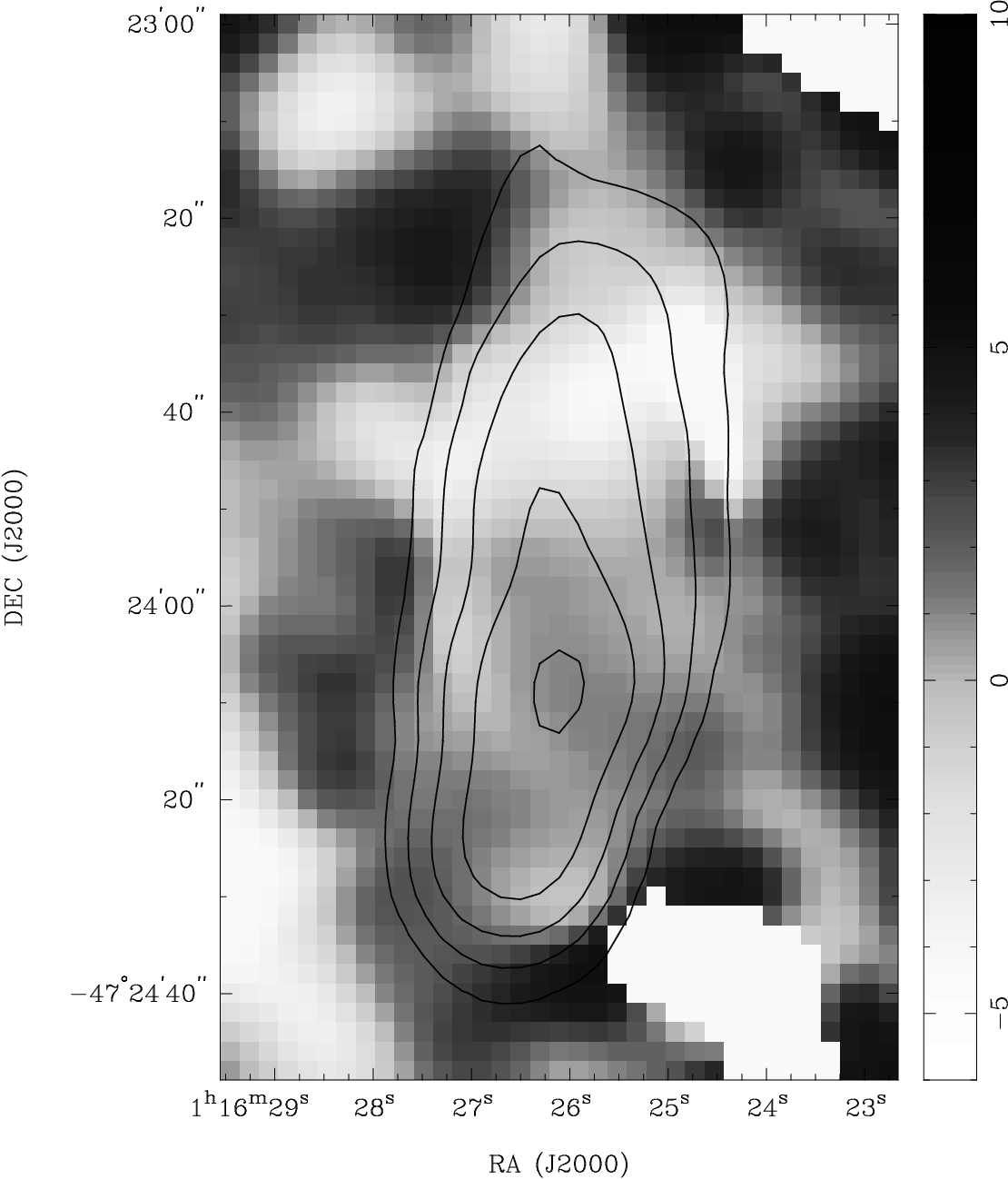}
\caption{
	RM distribution in the vicinity of the southern inner lobe S2.
	The RM, at a resolution of 12 arcsec FWHM, is shown using grey 
	scales in the range -6 to 10 rad m$^{-2}$.  Contours of a total 
        intensity image made using a beam
	of $9.5 \times 7.8$ arcsec$^{2}$ at 9$^{\circ}$ p.a. are
	overlayed; contours are at 1, 2, 4, and 8 mJy beam$^{-1}$.
	\label{fig9}
	}
\end{figure}

\clearpage

\begin{figure}
\epsscale{1.0}
\plotone{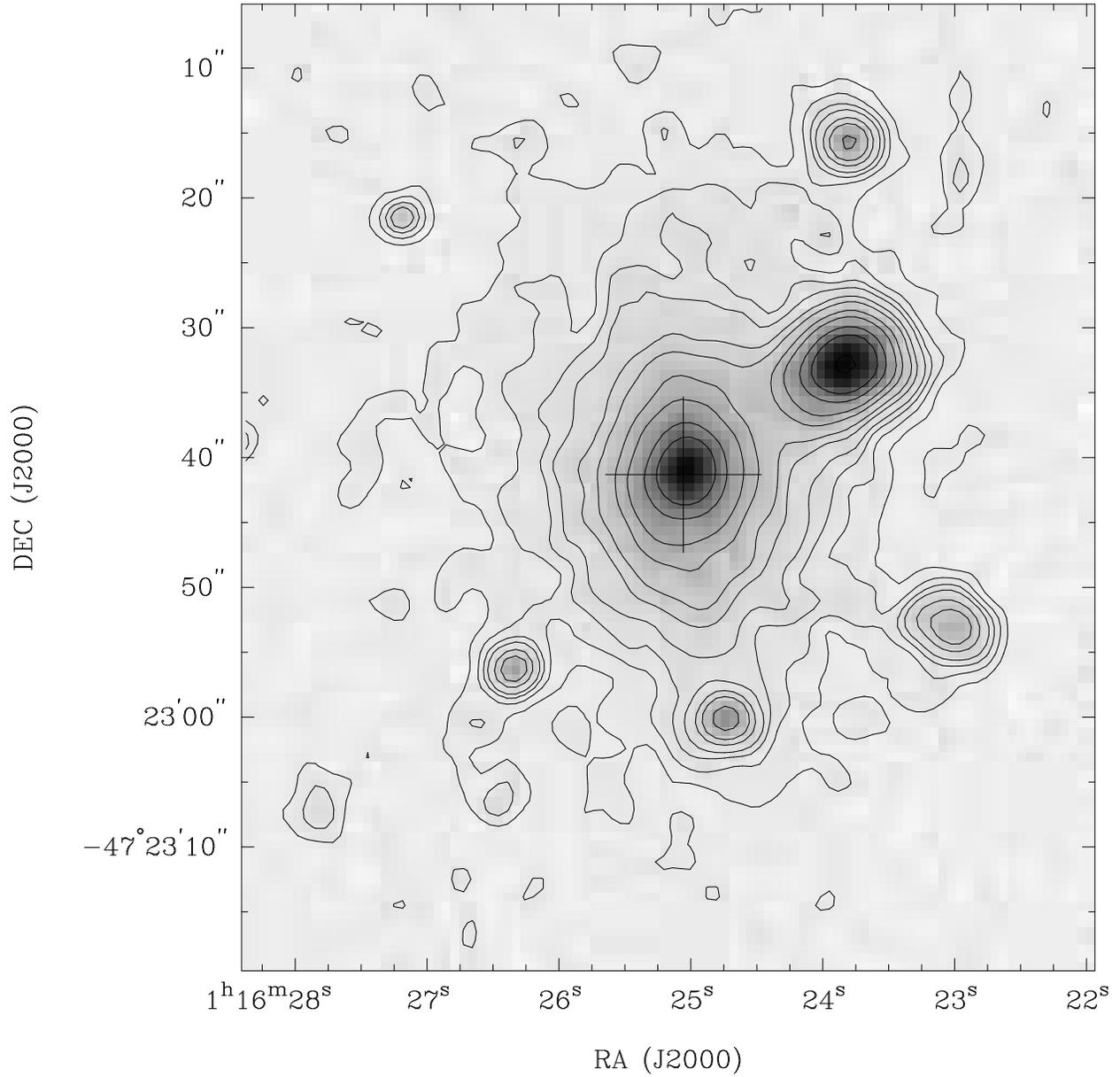}
\caption{
	The optical ID for J0116$-$473. The red UKST image 
	is displayed using grey-scales.  A smoothed version  obtained by 
	convolving with a 2-arcsec Gaussian is shown using contours;
	contours are at 2, 4, 6, 8, 12, 16, 
	24, 32, 48, 64, 96 per cent of the peak. The cross indicates
	the location of the radio core at 2496 MHz. 
	\label{fig10}
	}
\end{figure}

\end{document}